\documentclass{article}

\usepackage[preprint,nonatbib]{neurips_2020}

\usepackage[utf8]{inputenc} 
\usepackage[T1]{fontenc}
\usepackage{hyperref}
\usepackage{url} 
\usepackage{booktabs} 
\usepackage{amsfonts}
\usepackage{nicefrac}   
\usepackage{microtype} 
\usepackage{booktabs} 
\usepackage{microtype}
\usepackage{graphicx}
\usepackage{subfigure}
\usepackage{microtype}
\usepackage{hyperref}
\usepackage{url}
\usepackage{amsthm}
\usepackage{amsfonts}
\usepackage{color}
\usepackage{multicol}
\usepackage{comment}
\usepackage{lipsum}
\usepackage{afterpage}
\usepackage{amsmath}
\usepackage{mathrsfs}
\usepackage{algorithm}
\usepackage[noend]{algorithmic}

\title{Accurate Prediction of Free Solvation Energy of Organic Molecules via Graph Attention Network and Message Passing Neural Network from Pairwise Atomistic Interactions}

\author{%
  Ramin Ansari\thanks{Corresponding Author.} \\
  Department of Chemical Engineering\\
  University of Michigan\\
  Ann Arbor, MI, USA \\
  \texttt{raminans@umich.edu} \\
\And
  Amirata Ghorbani\\
  Department of Electrical Engineering\\
  Stanford University\\
  Stanford, CA, USA\\
  \texttt{amiratag@stanford.edu} \\
}

\begin{document}

\maketitle

\begin{abstract}
Deep learning based methods have been widely applied to predict various kinds of molecular properties in the pharmaceutical industry with increasingly more success.  Solvation free energy is an important index in the field of organic synthesis, medicinal chemistry, drug delivery, and biological processes.  However, accurate solvation free energy determination is a time-consuming experimental process. Furthermore, it could be useful to assess solvation free energy in the absence of a physical sample.  In this study, we propose two novel models for the problem of free solvation energy predictions, based on the Graph Neural Network (GNN) architectures: Message Passing Neural Network (MPNN) and Graph Attention Network (GAT).  GNNs are capable of summarizing the predictive information of a molecule as low-dimensional features directly from its graph structure without relying on an extensive amount of intra-molecular descriptors. As a result, these models are capable of making accurate predictions of the molecular properties without the time consuming process of running an experiment on each molecule.  We show that our proposed models outperform all quantum mechanical and molecular dynamics methods in addition to existing alternative machine learning based approaches in the task of solvation free energy prediction.  We believe such promising predictive models will be applicable to enhancing the efficiency of the screening of drug molecules and be a useful tool to promote the development of molecular pharmaceutics.

\end{abstract}

\section{Introduction}

The ability to predict physiochemical and biological properties of organic compounds is of significant importance in developing new materials with specific desired properties.  It allows us to compute properties without time-consuming and costly experiments as well as to discover novel compounds with extraordinary characteristics.  Common strategies to predict such properties are ab initio quantum mechanical simulations like Hartree-Fock (HF) \cite{echenique2007mathematical}, density functional theory (DFT) \cite{parr1980density, burke2012perspective}, molecular dynamics (MD) \cite{payne1986molecular, paquet2015molecular}, and hybrid quantum mechanics/molecular mechanics (QM/MM) \cite{gao1998combined, ratcliff2017challenges}.  Traditionally, these methods have been widely used to calculate various molecular properties without experimentation.  High-performance computational screening based on these approaches has become routine, giving researchers the ability to simulate the properties of thousands of chemical compounds as part of a single study.  However, their applications in real practice may be limited considering the high computational cost of running these advanced methods, as well as their low accuracy, especially for large systems \cite{wodrich2007accurate}. As a result, the use of machine learning models has become more widespread as they are computationally efficient and more accurate.

Recent developments in machine learning (ML), specifically deep learning (DL), have enabled researchers to predict the structure and properties of complex materials with an accuracy comparable to computational chemistry methods \cite{chibani2020machine, rupp2012fast}.  An ML algorithm trains a machine learning predictive model using a training set of input-output pairs so that the model is able to predict the output given the input. Deep learning models have enabled the use of large datasets where instead of using hand-crafted features, the models are able to learn hidden patterns in the data by itself.  In materials science, machine learning is critical in areas such as new material discovery and material property prediction \cite{butler2018machine, sanchez2018inverse}.  Computational chemistry simulations and experimental measurements are two conventional methods that are widely adopted in the field of materials science.  However, it is difficult to use these two methods to accelerate materials discovery and design because they are often time consuming and inefficient.  With the launch of the Materials Genome Initiative, a large effort has been made by material scientists to collect extensive datasets of materials properties.  These datasets include pre-existing computational simulations and experimental measurements.  Therefore, machine learning techniques can be used for finding patterns in this high-dimensional data, providing a fast and reliable solution for predicting the inherent properties of a large number of candidate molecules.  In recent years, ML algorithms have been widely used for material property prediction.  Also, they are often used in combination with quantum chemistry simulations such as DFT and MD for applications such as excited state dynamics \cite{ramakrishnan2015electronic, westermayr2020machine, westermayr2020combining, gomez2016design, pronobis2018capturing}, solubility of organic molecules \cite{gao2020accurate, huuskonen2000estimation, lusci2013deep, delaney2004esol}, bandgap of inorganic compounds \cite{lee2016prediction}, and thermodynamic stability \cite{schmidt2017predicting}.

Using ab initio methods such as DFT or MD, molecules are represented as a set of Cartesian coordinates which correspond to the atomic positions in 3D space.  Therefore, we can obtain properties related to the geometry and energy of the molecular system by solving the Schrodinger equation or Newton's equation of motion.  However, for ML algorithms, representing the atomic positions in 3D space is not the only way to predict the molecular properties.  In recent years, scientists have come up with various molecular representations for material property prediction \cite{sanchez2018inverse}. Some of these different molecular representations are: (1) SMILES \cite{weininger1989smiles, lim2019delfos} that uses short strings to describe the structure of chemical species; (2) Extended Connectivity Fingerprints (ECFPs) \cite{duvenaud2015convolutional, rogers2010extended} that breaks molecules into local neighborhoods and hashes them into a bit vector of an specified size; (3) Electronic density \cite{parr1989density, hirn2017wavelet} which is the measure of the probability of an electron being present at an infinitesimal element of space; (4) Coulomb matrix \cite{rupp2012fast} which represents the forces between each pair of atoms; (5) 3D geometry which is similar to inputs for DFT and MD simulations; (6) Bag of bonds and fragments; (7) Chemical environments \cite{bartok2013representing}; and etc.  The ML algorithms provide the benefit of a variety of molecular representations, which give scientists the ability to select the best representation for their property of interest.  In addition to the references cited above, we recommend references \cite{david2020molecular, ertl2010molecular, mater2019deep} for readers interested in these topics.

Molecules can also be expressed in the form of graphs, with the inclusion of the chemical information.  Graphs provide a natural way of describing molecular structures as atoms can be represented by nodes and chemical bonds by edges.  The basic chemical information can be encoded in the molecular graphs as the input.  The chemical information includes atomic features such as atom type and ionic radius, bond features such bond type and conjugation, and molecular features such as polarity and molecular weight.  Then, a graph based DL model such as GNN can be used to train a model. Graph theoretical approaches have been widely used to analyze chemical compounds \cite{zhou2018graph}, crystal structures, and even for the representation of reactions \cite{coley2019graph}.

Solvation free energy is the free energy change associated with the transfer of a solute molecule from ideal gas to a solvent at specific pressure and temperature. Solvation free energy, which describes the interaction between a solvent and dissolved molecules, is an important index in the fields of organic synthesis \cite{reichardt2011solvents, arnett1965solvent}, medicinal chemistry \cite{lipinski1997experimental, alhalaweh2012ph}, drug delivery \cite{steed2013supramolecular}, electrochemical redox reactions \cite{kim2019biological, takeda2016electron}, biological processes such as DNA and protein interaction \cite{chremos2018polyelectrolyte}, protein folding \cite{pace1996forces}, electronic and vibrational properties of biomolecules \cite{mashaghi2012hydration, mashaghi2012interfacial}, and etc.  Accurate solvation free energy determination is a time-consuming and costly process.  It would be particularly useful to assess solvation free energy in the absence of a physical sample as some molecules can be difficult or time-consuming to synthesize. ML models can predict multiple molecular properties prior to synthesis, avoiding the process entirely for molecules that do not demonstrate the desired properties. 

Recent studies demonstrated that ML models can successfully predict the solvation free energy of solutes in aqueous solutions (aka hydration free energy) \cite{delaney2004esol, wu2018moleculenet, lusci2013deep}.  Most of these ML models show an accuracy better than or comparable with ab initio simulations. ML models also have much faster calculation than computer simulations. While computational chemistry simulations can take up to a day for a single calculation, ML models can predict the solvation free energy of thousands of molecules in a few hours using a personal computer.  Unfortunately, the majority of ML models for solvation have been limited to aqueous solutions.

In the present work, we introduce a novel approach for predicting solvation free energy of solvent-solute pairs. Herein, we have used the graph based models: Message Passing Neural Network (MPNN) and Graph Attention Network (GAT). The results demonstrate that our graph-based models outperform the descriptor-based models and other graph based models in the literature in terms of prediction accuracy and computational efficiency.  The novelty of our model is that we take pair-wise interaction of solute and solvent into consideration.  We believe this methodology can be applied in other molecular property prediction beyond solvation free energy.

The rest of the present paper is outlined as follows: section 2 describes the embedding method for graph representation of molecules, a brief introduction on how MPNN and GAT models work, and feature aggregation.  In section 3, the model architecture, our results, and the performance of each approach are demonstrated in detail.  In section 4, we discuss the novelty of our models, their sources of error, and compare their performance with both MD and DFT simulation, as well as other ML models.  In the last section, we conclude our work and discuss directions for future work.

\section{Methods}

\subsection{Graph representation of molecules}
In graph neural networks, the first stage is to encode the molecule as a graph. In graph representation of molecule, the atoms and bonds that make up a molecule are mapped into sets of nodes and edges.  The graph is then augmented by the given information of the molecule which is stored as node (atom) features and edge (bond) features. 

\paragraph{Node Features} We use 8 features for each node where, after converting categorical features to one-hot vectors, each node is represented by a $31$-dimensional vector.  These 8 features are:

\begin{enumerate}
    \item The type of the atom is encoded as a 10-dimensional one-hot vector where the 10 choices are: Carbon, Nitrogen, Oxygen, Fluorine. Phosphorus, Sulfur, Clorine. Bromine. Iodine, or other atoms.
    \item The formal charge of the atom represented with an integer.
    \item Hybridization of the atom encoded as a 3-dimensional one-hot vector with categories “sp,” “sp2,” and “sp3.”
    \item Hydrogen bonding: A one-hot vector of whether this atom is a hydrogen bond donor or acceptor.
    \item Aromatic: A one-hot vector of whether the atom belongs to an aromatic ring.
    \item Degree: A one-hot vector of the degree (0-5) of the atom. Degree is the number of bonded neighbors.
    \item Number of Hydrogens: A one-hot vector of the number of hydrogens (0-4) to which the atom is connected.
    \item Calculated Partial charge.
\end{enumerate}

\paragraph{Edge features} Each edge between the nodes in the graph contains 4 features.  After converting the categorical features to one-hot vectors, each edge is represented by a $11$-dimensional vector.  These 4 features are:

\begin{enumerate}
    \item Bond type: A one-hot vector of the bond type, “single,” “double,” “triple," or “aromatic.”
    \item Same ring: A one-hot vector of whether the atoms in the pair are in the same ring.
    \item Conjugated: A one-hot vector of whether the bond is conjugated or not.
    \item Stereo: A one-hot vector of the stereo configuration of a bond.
\end{enumerate}

\begin{figure}[ht]
  \makebox[\textwidth][c]{\includegraphics[width=1.2\textwidth]{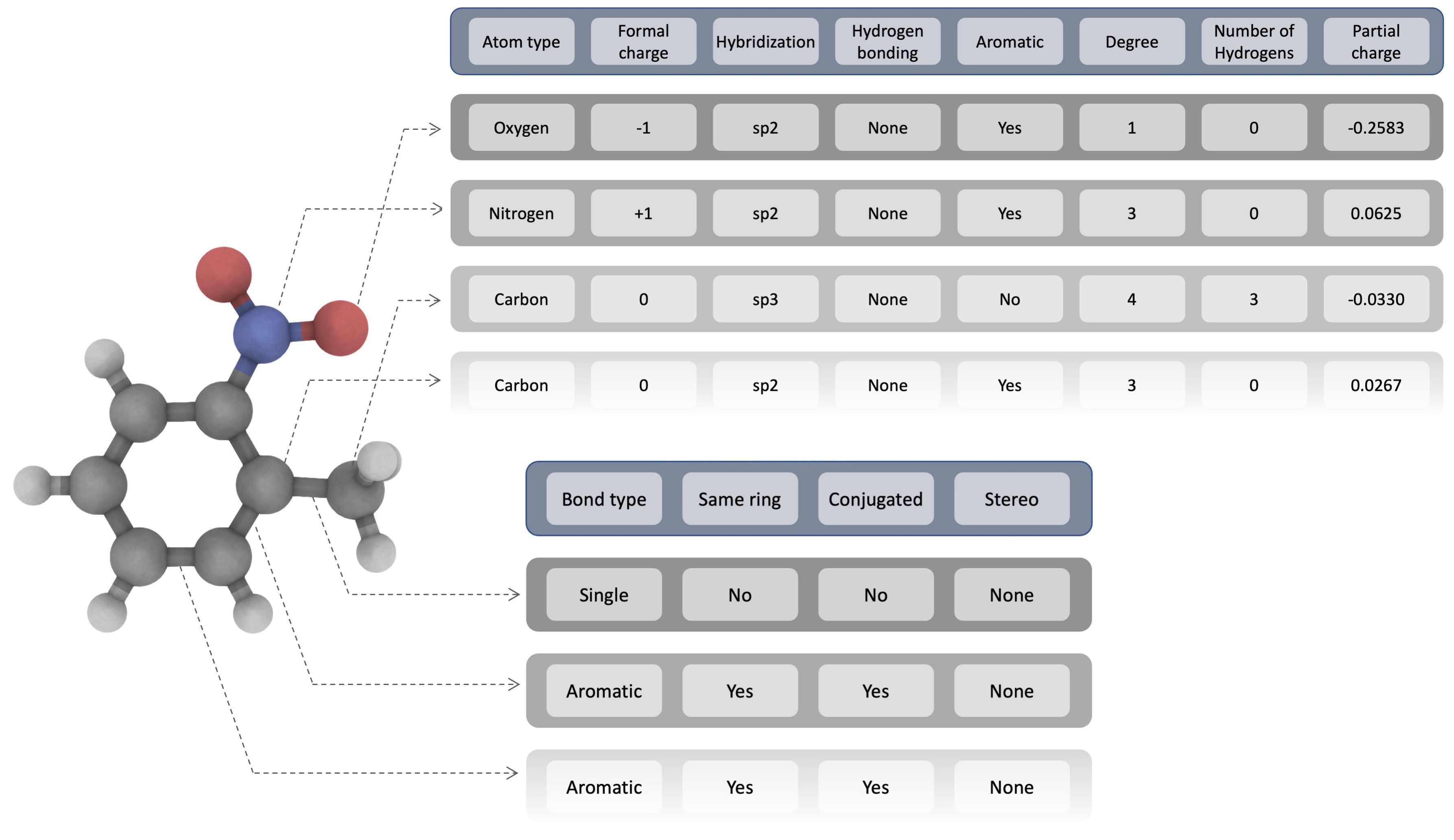}}%
  \caption{Example of node (atom) and edge (bond) features for a sample molecule, 2-Nitrotoluene.}
  \label{fig:features}
\end{figure}

In Figure \ref{fig:features} we have shown the molecule 2-Nitrotoluene as an example. It can be noted that the features are either numerical or categorical. The categorical features are encoded as one-hot vectors e.g. if there are five possible categories and the atom (or edge) is of category three, that feature is represented by a $5$-dimensional vector where the third element is one and the rest is zero. 

Once the molecule is transformed into its graph representation, it is fed into the encoder layer of the model. As we detail below, we use two different encoder models: MPNN and GAT. Each of these models take the graph object as input and output a learned representation of it such that the information of different atoms and edges are aggregated.

\subsection{Message Passing Neural Network (MPNN)}
In this work, we utilize and train an MPNN \cite{wu2020comprehensive, gilmer2017neural} architecture to generate vectors representing solvent and solute molecules.  In this section we will describe the MPNN framework in more detail.  Let G be the graph representation of a molecule (solute or solvent), where for each node $v$ we have its feature vector $h^0_v$, and for each edge $w$ we have its feature vector $e_{vw}$. The MPNN model goes through a number of \emph{message passing steps} where at each step $t$, the node $v$'s feature vector is updated to $h^t_v$. The update rule is:
\[m^{t+1}_v = \sum_{w \in N(v)} M_t(h^t_v, h^t_w, e_{vw})\]
\[h^{t+1}_v = U_t(h^t_v, m^{t+1}_v)\]
where N(v) denotes to the $v$'s neighboring nodes in $G$, and \(M_t\) and \(U_t\) are arbitrary functions that depends on hidden states and edges of the neighbouring nodes.   For readers interested in further reading on this we recommend references cited above.

As mentioned above, the size $h_v^0$ vectors in our problem is 31 (the number of the features after converting categorical features to one-hot vectors). However, the size of the feature vectors after message passing phase is set to be 64.  This is a hyperparameter that can be tuned.  The message passing phase runs for T=6 time steps which is defined as the number of times we update the output vector representation \(h^t_v\) for each node. 

\subsection{Graph Attention Network (GAT)}
Introduced in \cite{velickovic2018graph}, the GAT model is made of graph attention layers. A graph attention layer maps a set of node features to a new set of node features using a soft-attention mechanism. In short, for a given node, the graph attention layer outputs a weighted average of it and its neighbors' features where the weights are not fixed and are input-dependent. More specifically, for a given node $v$ and its feature-vector $h_v \in \mathbb{R}^F$, the graph attention mechanism outputs $h_v^\prime \in \mathbb{R}^{F^\prime}$ where:
$$
h_v^\prime = \sum_{w \in N(v)} \alpha_{vw} W h_v
$$
where $W \in \mathbb{R^{F^\prime \times F }}$ is a projection matrix and $F^\prime$ is the dimension of hidden features. In many cases, a non-linearity is also applied to the linear combination. The attention mechanism works such that the weights $\alpha_{vw}$ are determined based on the graph features. To make sure the weights are normalized, they are computed using a softmax formula:
$$
\alpha_{vi} = \frac{e^{l(a^T [Wh_v || Wh_i])}}{\sum_{w \in N(v)} e^{l(a^T [Wh_v || Wh_w])}}
$$
where $l$ is an activation function (we use the LeakyReLU function), $||$ stands for concatenation, and $a$ is a parameter vector in $R^{2F}$. The extension of a graph attention layer to multi-head attention is performed by using several projection matrices and then concatenating the resulting hidden features.

Looking at the above formulas, it is clear that the attention mechanism adds a large amount of flexibility to the model architecture; unlike graph convolutional layers, the attention layer is capable of assigning different levels of importance to each of the neighboring nodes in a dynamic fashion. Additionally, it adds interpretability to the model as one can investigate the neighbors that get the highest weight. In our experiments, we use a GAT model with two attention layers for the solvent and solute encoders with $F^\prime=42$ for both layers of the solvent encoder and $F^\prime=12$ for both layers of the solute model. We use multi-headed attention with $4$ heads in all layers of both models.

\subsection{Feature Aggregation}

After passing the molecule graph through GAT or MPNN models, the encoded output is a matrix with a size of [$\mbox{number of nodes}\times64$].  The main problem is that different molecules will be encoded into representations of different sizes, since different molecules have different number of nodes.  To be able to feed this representation into existing neural network architectures (e.g. feed-forward layer), it is necessary to be able to aggregate the information in the encoded graph representations of different size into a feature-vector with a fixed size.  To correct this, we use the set2set by Vinyals et al. \cite{vinyals2015order}, and weightedsum\&max (as implemented in DGL package \cite{wang2019deep}) readout functions for MPNN and GAT models, respectively.  Given the encoded graph, the set2set (or weightedsum\&max) layer computes a feature-vector for the whole graph using a readout function R:
\[\hat y = R(\{h^T_v | v \in G\}).\]

Further using the MPNN (or GAT) model, the input graph is encoded into a graph represented by a matrix with a size of [$\mbox{number of nodes}\times64$]. The set2set (or weightedsum\&max) layer maps this matrix into a vector size of 128. 

\begin{figure}[ht]
  \makebox[\textwidth][c]{\includegraphics[width=1.2\textwidth]{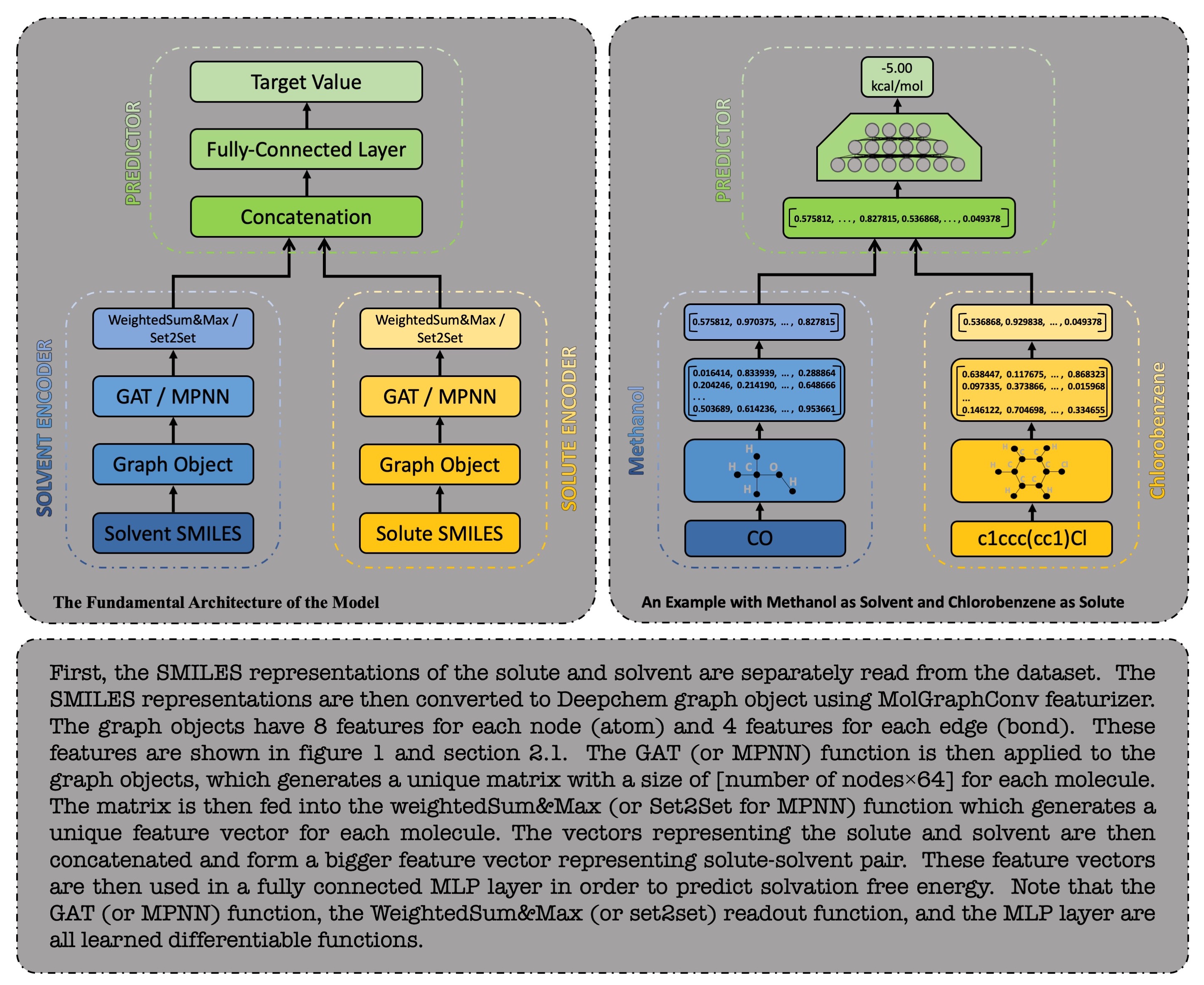}}%
  \caption{The fundamental architecture of the model (top left), an example of the model with methanol as the solvent and chlorobenzene as the solute (top right), and an explanation describing how the model works (bottom).}
  \label{fig:architecture}
\end{figure}

The final step is to aggregate the information of the solvent and solute molecules and make a prediction. This is done by simply concatenating the two molecules' feature vectors and then passing it through an MLP layer. The MLP layer then outputs the predicted solvation energy. The overall flow of the model is shown in Figure \ref{fig:architecture}.  Note that the message function \(M_t\), update function \(U_t\), graph attention mechanism, and the readout functions are all learned differentiable functions.

\section{Experiments}
\subsection{Computational Setup and Results}

In this study, 5,952 data points for non-aqueous solvents are collected with the Solv@TUM database version 1.0 \cite{hille2019generalized, stolov2017enthalpies, sedov2018abraham, mediatum1452571}.  This dataset contains a collection of experimentally measured partition coefficients for a large number of molecular solutes in non-aqueous solvents. There are 657 unique organic solvents and 146 unique organic solutes.  The partition coefficients are then converted to solvation free energies in kcal/mol.  Finally, the SMILES representation of 5,952 solute-solvent pairs and their solvation free energy in kcal/mol are prepared for the machine learning input.  Because the Solv@TUM dataset only contains common names of the solvents, an automated search is performed using PubChemPy \cite{pubchempy2014} library in order to obtain the SMILES strings of the compounds.  Less than 1\% of the compound names are not valid in PubChem database, therefore their SMILES strings are manually added to the dataset.  

For implementation of the neural network, we use PyTorch framework.  At the very first stage (Figure \ref{fig:architecture}), the SMILES representations are converted to a Deepchem graph object using the MolGraphConv \cite{kearnes2016molecular} featurizer as implemented in Deepchem \cite{leswing2019deep}. The graph objects have 8 features for each node (atom) and 4 features for each edge (bond).  The MPNN (or GAT) function is then applied to the graph objects, which generates a unique matrix for each molecule.  The resulting matrix is then fed into a readout phase which computes a feature vector for the whole graph using a given readout function (MPNN uses set2set and GAT uses WeightedSumAndMax).  Then we concatenate the vectors representing each solute and solvent pair.  The resulting vector is then used in a multilayer perception (MLP) model to predict the solvation free energy.  The MLP consists of 3 hidden layers each with 256, 256, and 128 hidden neurons, respectively. 

The dataset is split into train, validation, and test sets in an 8:1:1 ratio.  The train set has been used for training the models, while the validation set is used to give an estimate of model accuracy while tuning the model’s hyperparameters.  We perform an extensive grid search for tuning hyperparameters: learning rates, learning algorithms, and dimensions of hidden layers.  We choose the stochastic gradient descent (SGD) algorithm with Nesterov momentum, whose learning rate is 0.02, and 0.005 and momentum is 0.9, and 0.9 for MPNN and GAT models, respectively.  We also added a cosine annealing scheduler \cite{loshchilov2016sgdr} to the optimizer.  The test set is then used to give an unbiased estimate of the accuracy of the final tuned model.  To minimize the variance of the test runs we take an average of results over 10 independent runs.

\begin{figure}[ht]
  \makebox[\textwidth][c]{\includegraphics[width=1.2\textwidth]{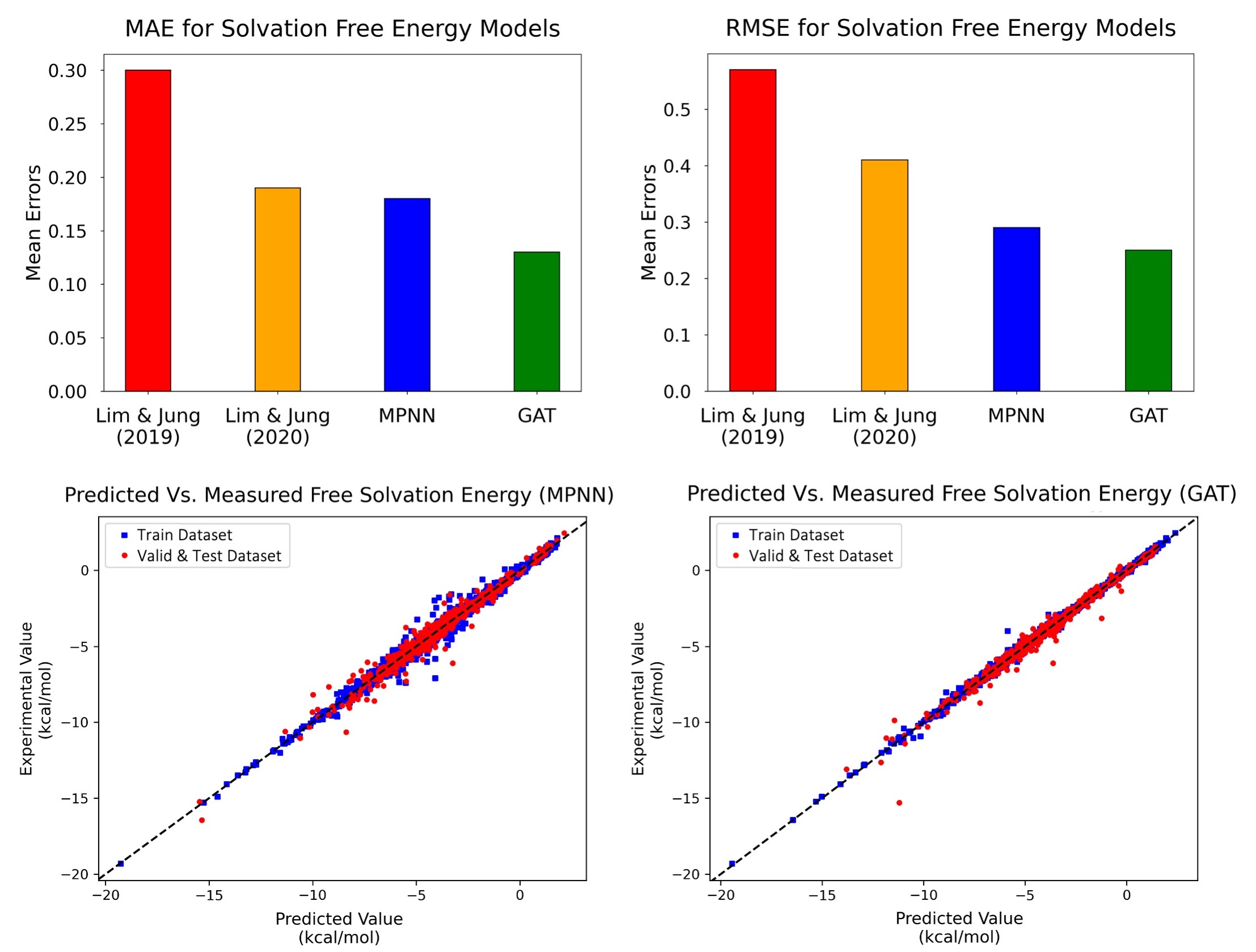}}%
  \caption{MAE (top left) and RMSE (top right) for Solvation free energy models.  MPNN and GAT are done in this study.  Scatter plot for true (y-axis) and ML predicted (x-axis) values of solvation energies for MPNN (bottom left), and GAT (bottom right).  All results are averaged over 10 independent runs.}
  \label{fig:pred_results}
\end{figure}

The performance of the two proposed models are shown in Figure \ref{fig:pred_results}.  For comparison, the performances of two other ML models proposed by Lim \& Jung \cite{lim2019delfos, lim2020mlsolv} are also shown in this figure.  We can see that both of our novel models perform better than the two example models proposed in the literature.  Moreover, from Figure \ref{fig:pred_results}, we can also see that the prediction errors for the MPNN model are 0.29 kcal/mol in RMSE and 0.18 kcal/mol in MAE, while results from the GAT model have an error of 0.25 kcal/mol in RMSE and 0.13 kcal/mol in MAE.  Together, these results indicate that the GAT architecture is the more suitable novel model for free solvation energy prediction.  Our GAT model has the lowest error obtained in any classical, quantum-mechanical, or ML based model for free solvation energy prediction.


\begin{figure}[ht]
  \makebox[\textwidth][c]{\includegraphics[width=1.2\textwidth]{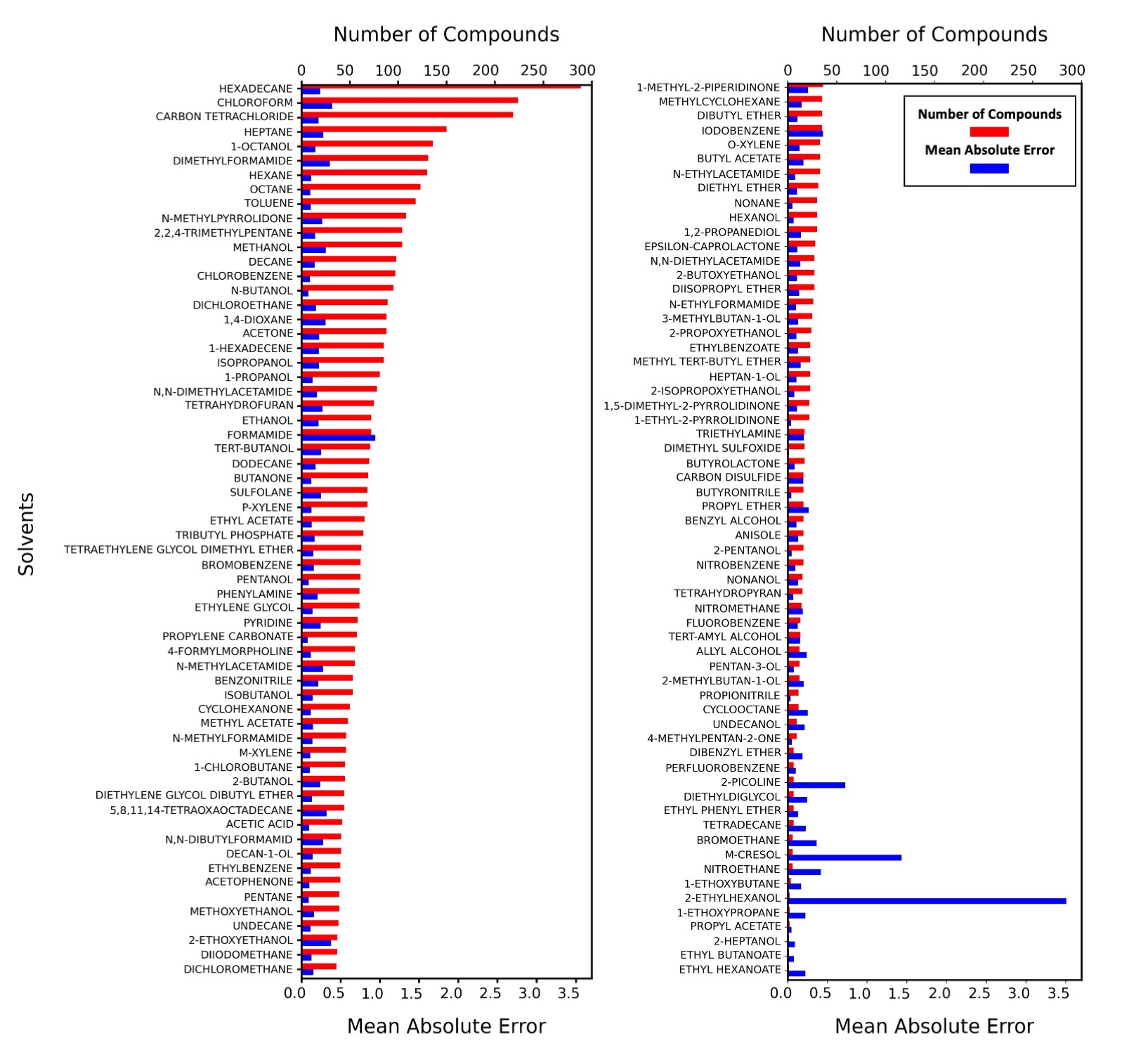}}%
  \caption{The mean absolute error and the number of compounds in the dataset for different solvents. The mean absolute error is calculated based on the prediction of the test dataset. The mean absolute error tends to be larger for the solvents that present less in the dataset.}
  \label{fig:solvent_errors}
\end{figure}

\section{Discussion}

In recent years, researchers have been widely used DFT \cite{wesolowski1994ab, zhao2011new}, MD \cite{shivakumar2010prediction}, and QM/MM \cite{zhang2018solvation} simulations to predict the solvation free energy of organic molecules in various solvents.  Despite the initial success of these methods, they are computationally expensive, often taking days to process one calculation. Additionally, they are complicated to set up even for experts; one needs to benchmark a variety of functional and basis sets (for DFT simulations), and force fields (for MD simulations) to find the best hyperparameters for the simulations.  Furthermore, these hyperparameters are usually different from molecule to molecule, meaning that new hyperparameters must be set for every molecular system.  Finally, these computational chemistry methods show a low accuracy with a mean absolute error (MAE) in the range of 0.6-1.0 kcal/mol \cite{schmidt2017predicting}.  However, our GAT and MPNN models are computationally cheaper and easier to set up with a much lower MAE of 0.13 and 0.18 kcal/mol, respectively.

Figure \ref{fig:solvent_errors} shows the mean absolute error and the number of compounds in the dataset for different solvents. The mean absolute error is calculated based on the prediction of the test dataset.  As it can be seen from this figure, the mean absolute error tends to be larger for the solvents that present less in the dataset. For example, the solvents 2-Ethylhexanol, M-cresol, and 2-Picoline only represent 2, 5, and 6 of the total number of compounds in the dataset and have an average absolute error of 3.51, 1.43, and 0.73 kcal/mol, respectively.  Clearly, the more data points that are available for each solvent, the lower the mean absolute error.  To lower the MAE for all solvents to more equitable levels, we should gather more extensive data for each solvent.

The Solv@TUM database contains a collection of experimentally measured solvation free energy of solutes in non-aqueous solvents. The data has been compiled from the works of at least 68 references.  It is apparent that each entry has been measured in somewhat different experimental conditions, with different equipment.  Moreover, each entry has been measured by different researchers and it is obvious that there might be some human error in the experimental values as well.  We believe that since our model is trained based on this dirty and coarse data, the small mean absolute error of 0.13 kcal/mol for the GAT model is somewhat inevitable.  Moreover, our model's error is in the range of the measurement errors of experimentally measured solvation free energies.  Thus, it is pretty impossible to have a lower prediction error.

A validation dataset is a sample of data held back from model training that is used to give an estimate of model accuracy while tuning the model’s hyperparameters.  The test dataset is also held back from the model training, but is instead used to give an unbiased estimate of the accuracy of the final tuned model.  There has been a trend in literature, especially in material property prediction, that researchers disregard validation datasets and use test datasets for both hyperparameter tuning and reporting the accuracy of the final tuned model.  We believe that for an unbiased model training, it is crucial for any ML study to have validation and test datasets.  Unlike the work done by Lim \& Jung \cite{lim2019delfos, lim2020mlsolv} for solvation free energy prediction and some other studies for material property prediction, we used the validation dataset for hyperparameter tuning and we reported the accuracy of the model using the test dataset.  We believe that this approach should also be employed by other researchers so that an unbiased estimate of the skill of the final tuned model could be computed.  

There exist, descriptor based ML models which rather than encoding the structure of a compound as a graph, instead the physio-chemical, structural, and electronic properties of compounds are separately calculated and manually added to the model \cite{jiang2021could}.  The Dragon software \cite{mauri2006dragon}, for example, can calculate 5,270 molecular descriptors.  These molecular descriptors are then used as input features to predict the property of interest.  However, having this many features may not be effective, as they increase the complexity of the model while it may not help much to improve the prediction accuracy.  Moreover, the descriptors could be strongly correlated; for example, adding atomic mass as an additional feature might not be useful since it is straightforwardly mapped to the atom types and thus already reflected in the atom type information.  Moreover, some of the features might not be related to the property of interest and negatively impact model performance.  For example, in the free solvation energy prediction features like "ring bridge count" might be totally irrelevant.  Therefore, researchers should extensively search and find those features which contribute most to the prediction variable in which we are interested in, while at the same time try to reduce overfitting, improve accuracy, and reduce training time.  This makes setting up the ML model so difficult and exclusive to the prediction output.  On the other hand, the graph molecular representation used in this work, only requires 8 and 4 features for each atom and bond, respectively.  The graph-based models are capable of generating and evaluating the physio-chemical, structural, and electronic properties of the compounds using this minimal number of features.  Using the graph structure and the given atom and bond features, these models directly compute a feature vector of desired size, which perfectly predicts the property of interest with a significantly low error.  This makes the ML setup so easy, fast, and, more importantly, flexible so that it can be used for any desired molecular property prediction.




\section{Conclusions}
In this study, two novel deep learning architectures, MPNN and GAT, are applied for free solvation energy predictions.  Each solute-solvent pair molecule is first converted to graph objects.  Each node (atom) has 8 atomic features: atom types, formal charges, hybridization, hydrogen bonding, aromaticity, degree, number of hydrogens, and partial charge. Each edge (bond) has 4 features: bond type, same ring, conjugation, and stereo configuration.  Then MPNN (or GAT) are applied to the solute and solvent pair separately.  The resulting vector feature representing solute and solvent are then concatenated and used in an MLP layer to predict the solvation free energy.  Our deep learning architectures shows better prediction than state of the art deep learning and quantum mechanical methods in terms of RMSE and MAE.  The MPNN model has an error of 0.29 kcal/mol in RMSE and 0.18 kcal/mol in MAE, while results from the GAT model show an error of 0.25 kcal/mol in RMSE and 0.13 kcal/mol in MAE.  This is the most accurate free solvation energy calculation in the literature to the best of our knowledge.

Most of the deep learning methods for solubility and solvation free energy only consider one solvent.  The novelty of our neural network model is that we take pair-wise interaction of solute and solvent into consideration.  We believe our architecture can be used for pair-wise interactions such as solvent-solute, protein-ligand, and etc.  Also, unlike chemical descriptor based models, our neural network architecture is able to find the physio-chemical and molecular properties needed for solvation free energy calculations with only a few atom and bond features.  This makes our model so flexible that can be used to learn and predict various molecular properties.  This methodology can be applied in other molecular property predictions beyond solvation free energy.   We believe such promising predictive models will be applicable to enhancing the efficiency of the screening of drug molecules, essentially as a useful tool to promote the development of molecular pharmaceutics.


\section{Conflicts of Interest}
There are no conflicts to declare. 
\section{Acknowledgments}
The author would like to thank Brianna T. Jackson for constructive proofreading of the manuscript.

\clearpage
\bibliography{references.bib}

\begin{thebibliography}{10}

\bibitem{echenique2007mathematical}
Pablo Echenique and Jos{\'e}~Luis Alonso.
\newblock A mathematical and computational review of hartree--fock scf methods
  in quantum chemistry.
\newblock {\em Molecular Physics}, 105(23-24):3057--3098, 2007.

\bibitem{parr1980density}
Robert~G Parr.
\newblock Density functional theory of atoms and molecules.
\newblock In {\em Horizons of quantum chemistry}, pages 5--15. Springer, 1980.

\bibitem{burke2012perspective}
Kieron Burke.
\newblock Perspective on density functional theory.
\newblock {\em The Journal of chemical physics}, 136(15):150901, 2012.

\bibitem{payne1986molecular}
MC~Payne, JD~Joannopoulos, DC~Allan, MP~Teter, and David~H Vanderbilt.
\newblock Molecular dynamics and ab initio total energy calculations.
\newblock {\em Physical review letters}, 56(24):2656, 1986.

\bibitem{paquet2015molecular}
Eric Paquet and Herna~L Viktor.
\newblock Molecular dynamics, monte carlo simulations, and langevin dynamics: a
  computational review.
\newblock {\em BioMed research international}, 2015, 2015.

\bibitem{gao1998combined}
Jiali Gao, Mark~A Thompson, et~al.
\newblock {\em Combined quantum mechanical and molecular mechanical methods},
  volume 712.
\newblock ACS Publications, 1998.

\bibitem{ratcliff2017challenges}
Laura~E Ratcliff, Stephan Mohr, Georg Huhs, Thierry Deutsch, Michel Masella,
  and Luigi Genovese.
\newblock Challenges in large scale quantum mechanical calculations.
\newblock {\em Wiley Interdisciplinary Reviews: Computational Molecular
  Science}, 7(1):e1290, 2017.

\bibitem{wodrich2007accurate}
Matthew~D Wodrich, Cl{\'e}mence Corminboeuf, Peter~R Schreiner, Andrey~A Fokin,
  and Paul von~Rague Schleyer.
\newblock How accurate are dft treatments of organic energies?
\newblock {\em Organic letters}, 9(10):1851--1854, 2007.

\bibitem{chibani2020machine}
Siwar Chibani and Fran{\c{c}}ois-Xavier Coudert.
\newblock Machine learning approaches for the prediction of materials
  properties.
\newblock {\em APL Materials}, 8(8):080701, 2020.

\bibitem{rupp2012fast}
Matthias Rupp, Alexandre Tkatchenko, Klaus-Robert M{\"u}ller, and O~Anatole
  Von~Lilienfeld.
\newblock Fast and accurate modeling of molecular atomization energies with
  machine learning.
\newblock {\em Physical review letters}, 108(5):058301, 2012.

\bibitem{butler2018machine}
Keith~T Butler, Daniel~W Davies, Hugh Cartwright, Olexandr Isayev, and Aron
  Walsh.
\newblock Machine learning for molecular and materials science.
\newblock {\em Nature}, 559(7715):547--555, 2018.

\bibitem{sanchez2018inverse}
Benjamin Sanchez-Lengeling and Al{\'a}n Aspuru-Guzik.
\newblock Inverse molecular design using machine learning: Generative models
  for matter engineering.
\newblock {\em Science}, 361(6400):360--365, 2018.

\bibitem{ramakrishnan2015electronic}
Raghunathan Ramakrishnan, Mia Hartmann, Enrico Tapavicza, and O~Anatole
  Von~Lilienfeld.
\newblock Electronic spectra from tddft and machine learning in chemical space.
\newblock {\em The Journal of chemical physics}, 143(8):084111, 2015.

\bibitem{westermayr2020machine}
Julia Westermayr and Philipp Marquetand.
\newblock Machine learning for electronically excited states of molecules.
\newblock {\em Chemical Reviews}, 2020.

\bibitem{westermayr2020combining}
Julia Westermayr, Michael Gastegger, and Philipp Marquetand.
\newblock Combining schnet and sharc: The schnarc machine learning approach for
  excited-state dynamics.
\newblock {\em The journal of physical chemistry letters}, 11(10):3828--3834,
  2020.

\bibitem{gomez2016design}
Rafael G{\'o}mez-Bombarelli, Jorge Aguilera-Iparraguirre, Timothy~D Hirzel,
  David Duvenaud, Dougal Maclaurin, Martin~A Blood-Forsythe, Hyun~Sik Chae,
  Markus Einzinger, Dong-Gwang Ha, Tony Wu, et~al.
\newblock Design of efficient molecular organic light-emitting diodes by a
  high-throughput virtual screening and experimental approach.
\newblock {\em Nature materials}, 15(10):1120--1127, 2016.

\bibitem{pronobis2018capturing}
Wiktor Pronobis, Kristof~T Sch{\"u}tt, Alexandre Tkatchenko, and Klaus-Robert
  M{\"u}ller.
\newblock Capturing intensive and extensive dft/tddft molecular properties with
  machine learning.
\newblock {\em The European Physical Journal B}, 91(8):1--6, 2018.

\bibitem{gao2020accurate}
Peng Gao, Jie Zhang, Yuzhu Sun, and Jianguo Yu.
\newblock Accurate predictions of aqueous solubility of drug molecules via the
  multilevel graph convolutional network (mgcn) and schnet architectures.
\newblock {\em Physical Chemistry Chemical Physics}, 22(41):23766--23772, 2020.

\bibitem{huuskonen2000estimation}
Jarmo Huuskonen.
\newblock Estimation of aqueous solubility for a diverse set of organic
  compounds based on molecular topology.
\newblock {\em Journal of Chemical Information and Computer Sciences},
  40(3):773--777, 2000.

\bibitem{lusci2013deep}
Alessandro Lusci, Gianluca Pollastri, and Pierre Baldi.
\newblock Deep architectures and deep learning in chemoinformatics: the
  prediction of aqueous solubility for drug-like molecules.
\newblock {\em Journal of chemical information and modeling}, 53(7):1563--1575,
  2013.

\bibitem{delaney2004esol}
John~S Delaney.
\newblock Esol: estimating aqueous solubility directly from molecular
  structure.
\newblock {\em Journal of chemical information and computer sciences},
  44(3):1000--1005, 2004.

\bibitem{lee2016prediction}
Joohwi Lee, Atsuto Seko, Kazuki Shitara, Keita Nakayama, and Isao Tanaka.
\newblock Prediction model of band gap for inorganic compounds by combination
  of density functional theory calculations and machine learning techniques.
\newblock {\em Physical Review B}, 93(11):115104, 2016.

\bibitem{schmidt2017predicting}
Jonathan Schmidt, Jingming Shi, Pedro Borlido, Liming Chen, Silvana Botti, and
  Miguel~AL Marques.
\newblock Predicting the thermodynamic stability of solids combining density
  functional theory and machine learning.
\newblock {\em Chemistry of Materials}, 29(12):5090--5103, 2017.

\bibitem{weininger1989smiles}
David Weininger, Arthur Weininger, and Joseph~L Weininger.
\newblock Smiles. 2. algorithm for generation of unique smiles notation.
\newblock {\em Journal of chemical information and computer sciences},
  29(2):97--101, 1989.

\bibitem{lim2019delfos}
Hyuntae Lim and YounJoon Jung.
\newblock Delfos: deep learning model for prediction of solvation free energies
  in generic organic solvents.
\newblock {\em Chemical science}, 10(36):8306--8315, 2019.

\bibitem{duvenaud2015convolutional}
David Duvenaud, Dougal Maclaurin, Jorge Aguilera-Iparraguirre, Rafael
  G{\'o}mez-Bombarelli, Timothy Hirzel, Al{\'a}n Aspuru-Guzik, and Ryan~P
  Adams.
\newblock Convolutional networks on graphs for learning molecular fingerprints.
\newblock {\em arXiv preprint arXiv:1509.09292}, 2015.

\bibitem{rogers2010extended}
David Rogers and Mathew Hahn.
\newblock Extended-connectivity fingerprints.
\newblock {\em Journal of chemical information and modeling}, 50(5):742--754,
  2010.

\bibitem{parr1989density}
RG~Parr and W~Yang.
\newblock Density-functional theory of atoms and molecules new york: Oxford
  univ, 1989.

\bibitem{hirn2017wavelet}
Matthew Hirn, St{\'e}phane Mallat, and Nicolas Poilvert.
\newblock Wavelet scattering regression of quantum chemical energies.
\newblock {\em Multiscale Modeling \& Simulation}, 15(2):827--863, 2017.

\bibitem{bartok2013representing}
Albert~P Bart{\'o}k, Risi Kondor, and G{\'a}bor Cs{\'a}nyi.
\newblock On representing chemical environments.
\newblock {\em Physical Review B}, 87(18):184115, 2013.

\bibitem{david2020molecular}
Laurianne David, Amol Thakkar, Roc{\'\i}o Mercado, and Ola Engkvist.
\newblock Molecular representations in ai-driven drug discovery: a review and
  practical guide.
\newblock {\em Journal of Cheminformatics}, 12(1):1--22, 2020.

\bibitem{ertl2010molecular}
Peter Ertl.
\newblock Molecular structure input on the web.
\newblock {\em Journal of cheminformatics}, 2(1):1--9, 2010.

\bibitem{mater2019deep}
Adam~C Mater and Michelle~L Coote.
\newblock Deep learning in chemistry.
\newblock {\em Journal of chemical information and modeling}, 59(6):2545--2559,
  2019.

\bibitem{zhou2018graph}
Jie Zhou, Ganqu Cui, Zhengyan Zhang, Cheng Yang, Zhiyuan Liu, Lifeng Wang,
  Changcheng Li, and Maosong Sun.
\newblock Graph neural networks: A review of methods and applications.
\newblock {\em arXiv preprint arXiv:1812.08434}, 2018.

\bibitem{coley2019graph}
Connor~W Coley, Wengong Jin, Luke Rogers, Timothy~F Jamison, Tommi~S Jaakkola,
  William~H Green, Regina Barzilay, and Klavs~F Jensen.
\newblock A graph-convolutional neural network model for the prediction of
  chemical reactivity.
\newblock {\em Chemical science}, 10(2):370--377, 2019.

\bibitem{reichardt2011solvents}
Christian Reichardt and Thomas Welton.
\newblock {\em Solvents and solvent effects in organic chemistry}.
\newblock John Wiley \& Sons, 2011.

\bibitem{arnett1965solvent}
Edward~M Arnett, WG~Bentrude, John~J Burke, and Peter~McC Duggleby.
\newblock Solvent effects in organic chemistry. v. molecules, ions, and
  transition states in aqueous ethanol1.
\newblock {\em Journal of the American Chemical Society}, 87(7):1541--1553,
  1965.

\bibitem{lipinski1997experimental}
Christopher~A Lipinski, Franco Lombardo, Beryl~W Dominy, and Paul~J Feeney.
\newblock Experimental and computational approaches to estimate solubility and
  permeability in drug discovery and development settings.
\newblock {\em Advanced drug delivery reviews}, 23(1-3):3--25, 1997.

\bibitem{alhalaweh2012ph}
Amjad Alhalaweh, Lilly Roy, Na{\'\i}r Rodr{\'\i}guez-Hornedo, and Sitaram~P
  Velaga.
\newblock ph-dependent solubility of indomethacin--saccharin and
  carbamazepine--saccharin cocrystals in aqueous media.
\newblock {\em Molecular pharmaceutics}, 9(9):2605--2612, 2012.

\bibitem{steed2013supramolecular}
Jonathan~W Steed and Jerry~L Atwood.
\newblock {\em Supramolecular chemistry}.
\newblock John Wiley \& Sons, 2013.

\bibitem{kim2019biological}
Jihyeon Kim, Sunghyun Ko, Chanwoo Noh, Heechan Kim, Sechan Lee, Dodam Kim,
  Hyeokjun Park, Giyun Kwon, Giyeong Son, Jong~Wan Ko, et~al.
\newblock Biological nicotinamide cofactor as a redox-active motif for
  reversible electrochemical energy storage.
\newblock {\em Angewandte Chemie International Edition}, 58(47):16764--16769,
  2019.

\bibitem{takeda2016electron}
Takashi Takeda, Ryosuke Taniki, Asuna Masuda, Itaru Honma, and Tomoyuki
  Akutagawa.
\newblock Electron-deficient anthraquinone derivatives as cathodic material for
  lithium ion batteries.
\newblock {\em Journal of Power Sources}, 328:228--234, 2016.

\bibitem{chremos2018polyelectrolyte}
Alexandros Chremos and Jack~F Douglas.
\newblock Polyelectrolyte association and solvation.
\newblock {\em The Journal of chemical physics}, 149(16):163305, 2018.

\bibitem{pace1996forces}
CN~Pace and BA~Shirley.
\newblock Forces contributing proteins of proteins.
\newblock {\em Faseb}, 10(1):75--83, 1996.

\bibitem{mashaghi2012hydration}
Alireza Mashaghi, Pouya Partovi-Azar, Tayebeh Jadidi, Nasser Nafari, Philipp
  Maass, M~Reza~Rahimi Tabar, Mischa Bonn, and Huib~J Bakker.
\newblock Hydration strongly affects the molecular and electronic structure of
  membrane phospholipids.
\newblock {\em The Journal of chemical physics}, 136(11):03B611, 2012.

\bibitem{mashaghi2012interfacial}
Alireza Mashaghi, P~Partovi-Azar, Tayebeh Jadidi, Nasser Nafari, Keivan
  Esfarjani, Philipp Maass, M~Reza~Rahimi Tabar, Huib~J Bakker, and Mischa
  Bonn.
\newblock Interfacial water facilitates energy transfer by inducing extended
  vibrations in membrane lipids.
\newblock {\em The Journal of Physical Chemistry B}, 116(22):6455--6460, 2012.

\bibitem{wu2018moleculenet}
Zhenqin Wu, Bharath Ramsundar, Evan~N Feinberg, Joseph Gomes, Caleb Geniesse,
  Aneesh~S Pappu, Karl Leswing, and Vijay Pande.
\newblock Moleculenet: a benchmark for molecular machine learning.
\newblock {\em Chemical science}, 9(2):513--530, 2018.

\bibitem{wu2020comprehensive}
Zonghan Wu, Shirui Pan, Fengwen Chen, Guodong Long, Chengqi Zhang, and S~Yu
  Philip.
\newblock A comprehensive survey on graph neural networks.
\newblock {\em IEEE transactions on neural networks and learning systems},
  2020.

\bibitem{gilmer2017neural}
Justin Gilmer, Samuel~S Schoenholz, Patrick~F Riley, Oriol Vinyals, and
  George~E Dahl.
\newblock Neural message passing for quantum chemistry.
\newblock In {\em International Conference on Machine Learning}, pages
  1263--1272. PMLR, 2017.

\bibitem{velickovic2018graph}
Petar Veli{\v{c}}kovi{\'{c}}, Guillem Cucurull, Arantxa Casanova, Adriana
  Romero, Pietro Li{\`{o}}, and Yoshua Bengio.
\newblock {Graph Attention Networks}.
\newblock {\em International Conference on Learning Representations}, 2018.

\bibitem{vinyals2015order}
Oriol Vinyals, Samy Bengio, and Manjunath Kudlur.
\newblock Order matters: Sequence to sequence for sets.
\newblock {\em arXiv preprint arXiv:1511.06391}, 2015.

\bibitem{wang2019deep}
Minjie Wang, Da~Zheng, Zihao Ye, Quan Gan, Mufei Li, Xiang Song, Jinjing Zhou,
  Chao Ma, Lingfan Yu, Yu~Gai, et~al.
\newblock Deep graph library: A graph-centric, highly-performant package for
  graph neural networks.
\newblock {\em arXiv preprint arXiv:1909.01315}, 2019.

\bibitem{hille2019generalized}
Christoph Hille, Stefan Ringe, Martin Deimel, Christian Kunkel, William~E
  Acree, Karsten Reuter, and Harald Oberhofer.
\newblock Generalized molecular solvation in non-aqueous solutions by a single
  parameter implicit solvation scheme.
\newblock {\em The Journal of chemical physics}, 150(4):041710, 2019.

\bibitem{stolov2017enthalpies}
Mikhail~A Stolov, Ksenia~V Zaitseva, Mikhail~A Varfolomeev, and William~E
  Acree.
\newblock Enthalpies of solution and enthalpies of solvation of organic solutes
  in ethylene glycol at 298.15 k: prediction and analysis of intermolecular
  interaction contributions.
\newblock {\em Thermochimica Acta}, 648:91--99, 2017.

\bibitem{sedov2018abraham}
Igor~A Sedov, Timur~M Salikov, Anisha Wadawadigi, Olivia Zha, Ellen Qian,
  William~E Acree~Jr, and Michael~H Abraham.
\newblock Abraham model correlations for describing the thermodynamic
  properties of solute transfer into pentyl acetate based on headspace
  chromatographic and solubility measurements.
\newblock {\em The Journal of Chemical Thermodynamics}, 124:133--140, 2018.

\bibitem{mediatum1452571}
Christoph~(1) Hille, Stefan~(2) Ringe, Martin~(1) Deimel, Christian~(1) Kunkel,
  William E.~(3) Acree, Karsten~(1) Reuter, and Harald~(1) Oberhofer.
\newblock Solv@tum v 1.0, 2018.

\bibitem{pubchempy2014}
Matt Swain, Eka~A. Kurniawan, Zach Pawers, Hsiao Yi, Leonardo Lazzaro, Björn
  Dahlgren, and Rickard Sjorgen.
\newblock Pubchempy.
\newblock {\em https:// github.com/mcs07/PubChemPy}, 2014.

\bibitem{kearnes2016molecular}
Steven Kearnes, Kevin McCloskey, Marc Berndl, Vijay Pande, and Patrick Riley.
\newblock Molecular graph convolutions: moving beyond fingerprints.
\newblock {\em Journal of computer-aided molecular design}, 30(8):595--608,
  2016.

\bibitem{leswing2019deep}
Bharath Ramsundar, Peter Eastman, Patrick Walters, Vijay Pande, Karl Leswing,
  and Zhenqin Wu.
\newblock Deep learning for the life sciences.
\newblock {\em OReilly: Sebastopol, CA, US}, 2019.

\bibitem{loshchilov2016sgdr}
Ilya Loshchilov and Frank Hutter.
\newblock Sgdr: Stochastic gradient descent with warm restarts.
\newblock {\em arXiv preprint arXiv:1608.03983}, 2016.

\bibitem{lim2020mlsolv}
Hyuntae Lim and YounJoon Jung.
\newblock Mlsolv-a: A novel machine learning-based prediction of solvation free
  energies from pairwise atomistic interactions.
\newblock {\em arXiv preprint arXiv:2005.06182}, 2020.

\bibitem{wesolowski1994ab}
T~Wesolowski and Arieh Warshel.
\newblock Ab initio free energy perturbation calculations of solvation free
  energy using the frozen density functional approach.
\newblock {\em The Journal of Physical Chemistry}, 98(20):5183--5187, 1994.

\bibitem{zhao2011new}
Shuangliang Zhao, Zhehui Jin, and Jianzhong Wu.
\newblock New theoretical method for rapid prediction of solvation free energy
  in water.
\newblock {\em The Journal of Physical Chemistry B}, 115(21):6971--6975, 2011.

\bibitem{shivakumar2010prediction}
Devleena Shivakumar, Joshua Williams, Yujie Wu, Wolfgang Damm, John Shelley,
  and Woody Sherman.
\newblock Prediction of absolute solvation free energies using molecular
  dynamics free energy perturbation and the opls force field.
\newblock {\em Journal of chemical theory and computation}, 6(5):1509--1519,
  2010.

\bibitem{zhang2018solvation}
Pan Zhang, Lin Shen, and Weitao Yang.
\newblock Solvation free energy calculations with quantum mechanics/molecular
  mechanics and machine learning models.
\newblock {\em The Journal of Physical Chemistry B}, 123(4):901--908, 2018.

\bibitem{jiang2021could}
Dejun Jiang, Zhenxing Wu, Chang-Yu Hsieh, Guangyong Chen, Ben Liao, Zhe Wang,
  Chao Shen, Dongsheng Cao, Jian Wu, and Tingjun Hou.
\newblock Could graph neural networks learn better molecular representation for
  drug discovery? a comparison study of descriptor-based and graph-based
  models.
\newblock {\em Journal of cheminformatics}, 13(1):1--23, 2021.

\bibitem{mauri2006dragon}
Andrea Mauri, Viviana Consonni, Manuela Pavan, and Roberto Todeschini.
\newblock Dragon software: An easy approach to molecular descriptor
  calculations.
\newblock {\em Match}, 56(2):237--248, 2006.

\end{thebibliography}
\bibliographystyle{unsrt}

\clearpage
\appendix
\end{document}